\documentclass[12pt,twoside,a4paper]{article}
\usepackage{imm17_4}
\usepackage[english]{babel} 
\usepackage[cp1251]{inputenc} 
\usepackage[T2A,OT1]{fontenc} 

\newcommand{\Leftclt}{A.\,V.\,Makarenko,\,A.\,V.\,Pravdivtsev}

\newcommand{\Rightclt}{Architectural solutions of conformal network-centric staring-sensor systems\ldots}

\newcommand{\UDK}{00.00.00}

\begin{document}

\Mainclt 

\begin{center}
\Large{\bf Architectural solutions of conformal network-centric staring-sensor systems with spherical field of view}\\[2ex]
\end{center}

\begin{center}
\large\bf{A.\,V.\,Makarenko\footnote{E-mail: avm.science@mail.ru},\,A.\,V.\,Pravdivtsev}\\[2ex]
\end{center}

\begin{center}
\normalsize{R~\&~D~Group~“Constructive Cybernetics”}
\\
\normalsize{P.~O.~Box, 560, Moscow, 101000, Russia}\\[3ex]
\end{center}

\begin{quote}\small
The article presents the concept of network-centric conformal electro-optical systems construction with spherical field of view. It discusses abstract passive distributed electro-optical systems with focal array detectors based on a group of moving objects distributed in space. The system performs conformal processing of information from sensor matrix in a single event coordinate-time field. Unequivocally the construction of the systems which satisfy the different criteria of optimality is very complicated and requires special approaches to their development and design. The paper briefly touches upon key questions (in the authors' opinion) in the synthesis of such systems that meet different criteria of optimality. The synthesis of such systems is discussed by authors with the systematic and synergy approaches.
\end{quote}


\begin{Keyworden}
EOS, network-centric systems, architecture, wide angle optical systems, MWIR.
\end{Keyworden}


\setcounter{equation}{0}
\setcounter{lem}{0}
\setcounter{teo}{0}



\section{Introduction}
\setcounter{equation}{0}
Modern electro-optical systems (EOS) with the focal plane array detectors (FPA), located on the mobile carriers have high detectivity and resolution of the “object/background” contrast and of the angular coordinates.  Due to the passive working principle of objects locating, these stations are highly concealed.  So such systems are one of the major information sources about the environment in some tasks.

Current trends in the construction of EOS with FPA detectors are based on a group of moving objects is transitioning from scanning systems to a constant view by a group of the sensor with specified observation area~--~staring-sensor systems. The sensors are distributed on the carrier surface to form discrete analog of conformal receiving antenna. The advantages of staring-sensor systems are: increased mechanical reliability, the lack of unobserved zones during the scan, the lack of image blurring caused by scanning, increased the “signal/background” ratio, a simplified scheme of image processing, and so on. But in cases of spherical or hemispherical fields of view such systems require a large number of the sensors, which leads to the increased cost. This disadvantage became negligible due to constant increasing of the detectors resolution (the resolution of the available detectors for main sub-bands from UV to LWIR is $1K\times1K$ pixels, with expectation for the near future of $2K\times2K$ pixels).

The next logical step in the development of passive staring-sensor EOS is to form a unit conformal receiving antenna by combining signals from multiple EOS from different carriers. These carriers coordinate to perform the task and are compactly situated in the space. It is allow to form the unity field of view for the whole group. This conception of conformal EOS construction fits in the paradigm of network-centric organization of sensor networks\cite{bib:article_Cebrowski_98} and allows to talk about the network-centric distributed EOS.

It is worth noting that with net-centric distributed EOS the authors are considering two interconnected aspects. In the first place the information from group's sensor array is carried out jointly into a single unit of event-coordinate-time field. This field is tied to observable objects, informational, management and executive agents, events, and the data itself. It also integrates diverse information from various different sources. Thus, algorithms for processing data from EOS sensors are operating at a high level of a priori and operational information support. In the second place, the information redundancy and unsteadiness of sensors array generates a high level of information completeness and accuracy of input data. Along with external data, it is possible to create a synergistic effect: reconstruction of missing or distorted information. As a result the entire set of properties generates the emergence of network-centric EOS.

Unequivocally the construction of the systems which satisfy the different criteria of optimality is very complicated and requires special approaches to their development and design. The paper briefly touches upon key questions (in the authors' opinion) in the EOS synthesis that meet different criteria of optimality. The synthesis of such systems is discussed by authors with the systematic and synergy approaches. The paper is organized as follows: Section~\ref{sect:architecture} discussed construction of network-centric architecture of the EOS; Section~\ref{sect:OS} is devoted to the main issues of wide-angle MWIR optical system synthesis. The conclusions of the work in general are in Sec.~\ref{sect:conclusion}.

\section{Architecture of Network-centric EOS}
\label{sect:architecture}
\setcounter{equation}{0}

\begin{figure}
 \begin{center}
 \includegraphics[width=156mm,height=95mm]{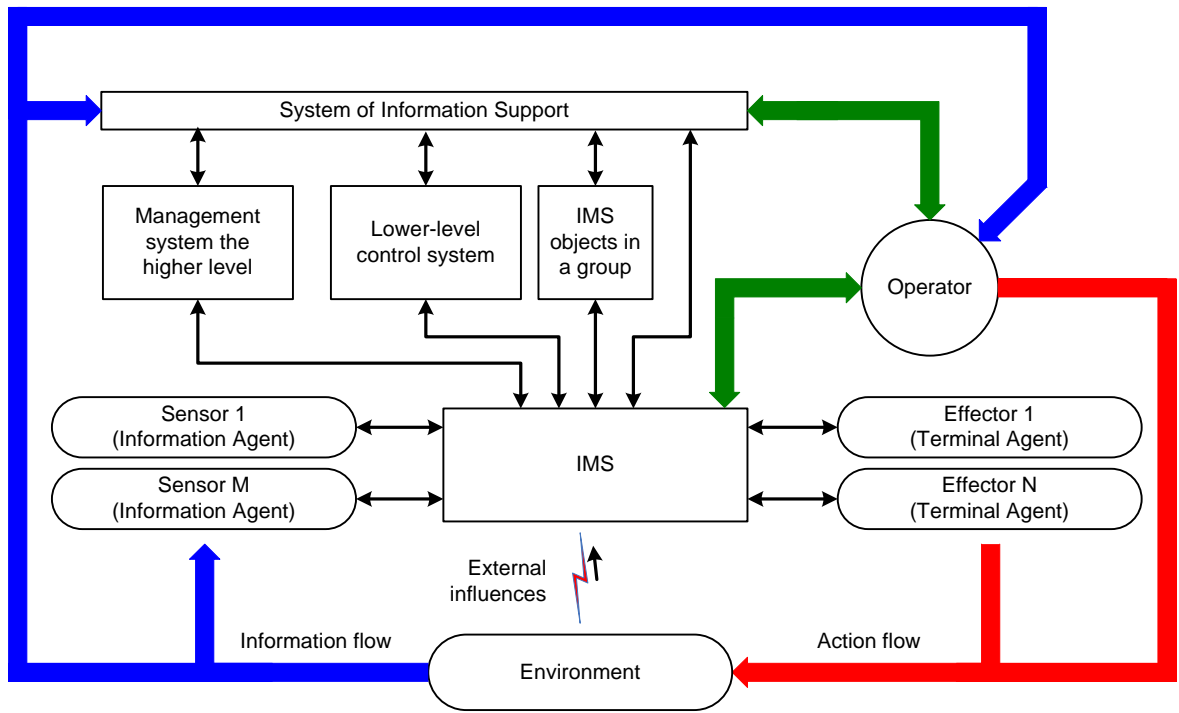}
 \end{center}
 \caption{The cybernetic paradigm of network-centric EOS.}\label{fig:ArchitSystem}
 \end{figure}

Using the cybernetic paradigm\cite{bib:book_Viner_61}, distributed network-centric EOS is in the first place a measurement and information system~--~a sensor (information agent). The cybernetic paradigm of interaction between EOS and external system and environment, adopted by the authors is shown in Fig.~\ref{fig:ArchitSystem}.

In the Fig.~\ref{fig:ArchitSystem} sensor means a set of information-measurement tools based on the carrier (robotic system) including a group of EOS, forming a conformal receiving antenna. The effector (terminal agents) means a set of tools for active influence, including maneuvers and tactical action of the carrier. The system of information support is a combination of internal and external tools for a priori and operational information support including mission, on-board geographical-information system and so on. Management system of the higher level is a management system which has hierarchy level higher than the one of this carrier. Information management system (IMS) of the group is a IMS of the other carriers which operates together and in concord with this carrier. Lower level control system is a control systems of the objects which have lower hierarchy level than this carrier.

Taking into account the heterogeneity of the paradigm (the system includes a human operator), its detail and formalization is conducted with the assistance of the synergetic provisions\cite{bib:book_Haken_04, bib:book_Haken_06}.

\begin{figure}
 \begin{center}
 \includegraphics[width=160mm,height=35mm]{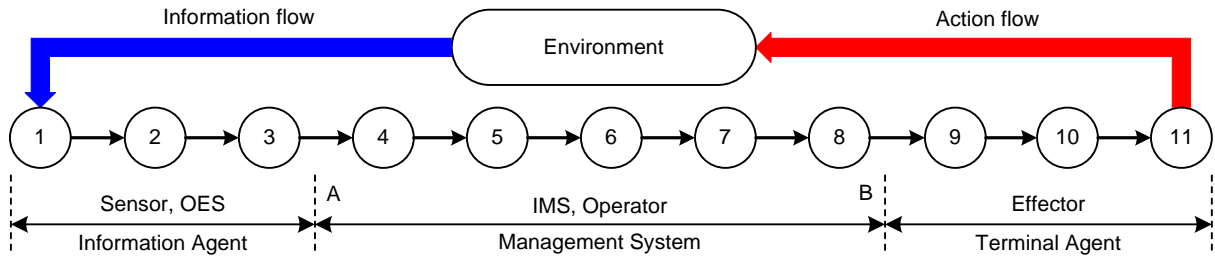}
 \end{center}
\caption{The stages of information processing in the system: 1~--~Acquisition and normalization of raw data; 2~--~Detection of the object; 3~--~The classification and diagnosis of the object; 4~--~Refinement object's coordinate and noncoordinate information; 5~--~Creation of operational and tactical situation model; 6~--~Recognition intentions the opposite side; 7~--~The prediction of the operational and tactical situation; 8~--~Synthesis of counteraction strategies; 9~--~The create of counteraction program; 10~--~Generation of counteraction commands; 11~--~Executing of counteraction commands.}
\label{fig:DataTransform}
\end{figure}

One of the major issues of network-centric EOS synthesis~--~is an optimal distribution of functions between the EOS and IMS. This approach is due to the ultrahigh information integration of the network-centric EOS in the global information management grid. A generalized scheme of information processing in the system is shown in Fig.~\ref{fig:DataTransform}.

Its necessary to note that the separation of steps between the Sensor and IMS~--~is the border “A”, and between IMS and Effector - border “B”, Fig.~\ref{fig:DataTransform} is an approximation. Variation of their actual locations leads to substantial differences in the architecture of carrier's information systems; requirements for “functional” and “intelligence” of sensors, IMS, effectors; parameters of the command/data exchange between the system's elements; characteristics of communication channels between the elements, capability of physical realizability of the system. Therefore, more precise definition of the actual position of the boundaries “A” and “B”, and as a consequence, network-centric architecture of the EOS, should be subject to optimization studies, with a given objective function and constraints (including economic and organizational). Direct genetic methods for global optimization, which support non-differentiable objective functions are used for this purpose. In authors' experience, the differential evolution method is well-proven\cite{bib:article_Storn_97, bib:book_Storn_05}.

EOS configuration for optimization is described by the following leading parameters: weight,  dimensions, power consumption; number of sensors; combination of spectral bands, the effective number of pixels per steradian; instantaneous field of view (IFOV) of the pixel, frame rate, sensitivity and noise immunity of the detector, the geometric factor (the range of distances to the observed object and range of distances between the objects of the group), the quality and stability of data processing and decision making algorithms, the composition and quality of required a priori and operational information supply; the performance of computing platform.

The solution of this optimization problem in one step is rather difficult, so it is logical to apply the iterative hierarchical method for the synthesis of complex systems which emphasizes comprehensive simulation and modeling technologies that operate in the “multiphysics” conception.

A special issue of synthesis and function of the described class of EOS~--~is the boundary configurations, where the distributed and network-centric properties are preserved. Degradation of network-centric systems in this case is determined by the following parameters: geometrical factor, the number of sensors, the composition and quality of the a priori and operational support.

\subsection{Logical-kinematical filtering of hypotheses}

A major disadvantage of the passive EOS is the inability to perform direct measurement or calculation of the slant range to the observed object (OO). It reduces the EOS information quality and leads to the difficulties in analyzing of the OO trajectories. However, as shown in several papers\cite{bib:report_Peli_02, bib:article_Cardillo_02}, trajectory analysis of signal responses observed by direction-finding station is extremely necessary for enhancing the information capabilities of robotic systems.

A way to restore the completeness of the coordinate data of the OO without constructing a direct finder triangle is being developed now. The network-centric EOS discussed in this paper is sufficiently adapted to this task. The approach is based on probabilistic-statistical methods of slant range estimation based on indirect measurements. The method takes into account the hypotheses of kinematic constraints on OO's trajectory.

Its necessary to note that most part of known approaches\cite{bib:report_Peli_02, bib:article_Daum_03, bib:report_Madyastha_05, bib:report_Tartakovsky_00} as a rule are reduced to a collection of isolated individual subtasks without good grounds. All available information cannot be fully used in the each of these subtasks. Additionally algorithms are usually built in a local coordinate system of the FPA or in the carrier coordinate system. Modified Kalman-Bucy filters are used, which assumes the solution is in the linear approximation. During the construction of feasible decisions area, rigid, a priori given inequalities are used; while the solution is sought in a class of parametric optimization.

In this paper, the authors state the task which extends this approach in terms of removing the described restrictions and enhancing its information and the probability-precision capabilities. The approach allows to recover the most probable trajectory of the object in the global coordinate system, to classify the object and diagnose its condition. The self-consistent nonlinear filter used to obtain these estimates is a feature of the proposed solution. Additionally classification/diagnosis of the object and its trajectory synthesis are carried out jointly. Kinematic trajectory evaluation of the object's coordinates; logical-probabilistic models of space and environment surrounding the object, the probabilistic and dynamic behavior of the object are combined in the whole. In addition, the algorithm makes a decision based on a priori and operational defined constraints and preferences provided in the form of non-rigid probability inequalities.

The authors suggest to considering a spatial (3-dimensional) rectangular coordinate system (R-CS). The position in space of a point~$P_W$ is defined by radius-vector~${\mathbf r}_W$. Additionally to the R-CS, a coordinate system associated with the Earth (E-CS), for example WGS-84~\cite{bib:report_WGS84_00}, is defined. Geodesic coordinates of a point~$P_W$ relative to the surface of base (terrestrial) ellipsoid of revolution~($S_{BE}$) are: $h^{gd}_W$~--~altitude, $\psi^{gd}_W$~--~latitude and $\phi^{gd}_W$~--~longitude. The surface~$S_{BE}$ is correlated with vector function~$\mathbf{b}_{BE}(\psi^{gd}_W,\,\phi^{gd}_W,\,t)$ which describe the map information (hereinafter referred to simply as “map”). Also defined is  the vector function~$\mathbf{c}_{BE}(h^{gd}_W,\,\psi^{gd}_W,\,\phi^{gd}_W,\,t)$ which describes weather and climatic properties of the atmosphere and parameters of propagation medium.

The carrier of passive EOS (mobile robotic system)~$\mathbf{V}$ and OO~$\mathbf{T}$ is represented by material points~$P_V$  and~$P_T$ respectively. Their positions in R-CS are defined by corresponding radius-vectors~$\mathbf{r}_V$ and~$\mathbf{r}_T$. Such assumption is correct if the distance between the OO, carrier and other obstacles is much greater than their characteristic dimensions.

The direct finder station is characterized by a vector of parameters~$\mathbf{b}_V|s_V$ in state~$s_V$ from the ensemble~$s_V\in\mathrm{S}_V\subset\mathbb{N}$,~$\mathrm{S}_V=\overline{1,\,S_V}$. The OO can belong to the class~$g_T$ from the set~$g_T\in\mathrm{G}_T\subset\mathbb{N}$,~$\mathrm{G}_T=\overline{1,\,G_T}$, and be in state~$s_T$ from the ensemble~$s_T\in\mathrm{S}_T|g_T\subset\mathbb{N}$, $\mathrm{S}_T|g_T=\overline{1,\,S_T|g_T}$. For every of~$s_T|g_T$  combinations is known the following vectors: $\mathbf{b}_T$~--~parameters of the object, $\mathbf{c}_T$~--~ways (tactic) of use.

Suppose that in $k$-th time~$t[k]$ EOS measures values:
 \begin{equation}\label{eq:r_d_observe}
 \tilde{\mathbf{r}}_V[k]=\mathbf{r}_V[k]+\mathbf{r}^\epsilon_V[k], \quad
 \tilde{\mathbf{d}}_{VT}[k]=\mathbf{d}_{VT}[k]+\mathbf{r}^\epsilon_{VT}[k],
 \end{equation}
where:~$\mathbf{d}_{VT}$~--~the direction vector of the line~$L_{VT}$~--~the line of sight of the OO; $\mathbf{r}^\epsilon_V$, $\mathbf {d}^\epsilon_{VT}$~--~the errors of measurement of relevant parameters. Simultaneously, there is fixation of values~$s_V[k]$. In the system there are also the following values: $\mathbf{c}_{TO}[k]$~--~vector of observation conditions; $\mathbf{s}_{TC}[k]$~--~state vector operational tactical situation.

A collection~$\mathrm{K}$ of consecutive measurements~$k\in\mathrm{K}\subset\mathbb{N}$, $\mathrm{K} = \overline{1, \, K}$ form a set:
 \begin{equation}\label{eq:collect_observe}
 OB|_\mathrm{K}=\left\lbrace
 \tilde{\mathbf{r}}_V|_\mathrm{K},\,
 \tilde{\mathbf{d}}_{VT}|_\mathrm{K},\,
 t|_\mathrm{K},\,
 s_V|_\mathrm{K},\,
 \mathbf{c}_{TO}|_\mathrm{K},\,
 \mathbf{s}_{TC}|_\mathrm{K}
 \right\rbrace.
 \end{equation}

A set~$\mathrm {N}$ of consecutive estimates~$n\in\mathrm{N}\subset\mathbb{N}$, $\mathrm{N}=\overline{1,\,N}$ form a set: $\hat{s}_T|_\mathrm{N}$, $\hat{\mathbf{Q}}_T|_\mathrm{N}$, in relation to the assessment~$\hat{g}_T$. The matrix~$\mathbf{Q}_T\lbrack n\rbrack$~--~is the kinematic parameters of the trajectory of the OO at time~$t\lbrack n \rbrack$:
 \begin{equation}\label{eq:cinematic matrix}
 \mathbf{Q}_T\lbrack n\rbrack=
 \left\lbrack{\mathbf{r}_T\lbrack n\rbrack\quad\dot{\mathbf{r}}_T\lbrack n\rbrack\quad\ddot{\mathbf{r}}_T\lbrack n\rbrack}\right\rbrack.
 \end{equation}

The followings interconsistency estimates needs to be get on the base of set~(\ref{eq:collect_observe}), for $K\geqslant 3$:
 \begin{equation}\label{eq:Alg_main}
 \begin{gathered}
 \left\lbrace{\hat{g}_T,\,\hat{s}_T|_\mathrm{N},\,\hat{\mathbf{Q}}_T|_\mathrm{N}}\right\rbrace=\\[2pt]
 \Alg\left\langle{ OB|_\mathrm{K},\,
                 t|_\mathrm{N},\,
                 \mathbf{b}_V|_{\mathrm{S}_V},\,
                 \mathbf{b}_T|^{\mathrm{G}_T}_{\mathrm{S}_T},\,
                 \mathbf{c}_T|^{\mathrm{G}_T}_{\mathrm{S}_T},\,
                 \mathbf{b}_{BE},\,
                 \mathbf{c}_{BE}}\right\rangle.
 \end{gathered}
 \end{equation}

Note that the problem~(\ref{eq:Alg_main}) is inherently~--~is a joint task of distinguishing and evaluating hypotheses by means of nonlinear filtering. Its solution ends steps of~3 and~4 of information processing in the system, see Fig.~\ref{fig:DataTransform}.

Depending on the ratio of time intervals~$T_\mathrm {K}$ and~$T_\mathrm{N} $ solution of the problem~(\ref{eq:Alg_main}) is split: on the approximation of the net, $T_\mathrm{N}\subseteq T_\mathrm{K}$; extrapolation, $T_\mathrm{N}\cap T_\mathrm{K}=\emptyset$; or a combination thereof. And: $T_\mathrm{K} = [\min t|_\mathrm{K}, \,\max t|_\mathrm {K}]$, $T_\mathrm{N} =[\min t |_\mathrm {N}, \, \max t|_\mathrm {N}]$. The configuration of the interval~$T_\mathrm {N}$ divides the problem~(\ref{eq:Alg_main}), by application into three classes:

{\bf A}~--~operative. The purpose of $\Alg$ is the evaluation of the instantaneous slant range to the object, the identification of its flexibility and its velocity vector prediction.

{\bf B}~--~tactical. The purpose of $\Alg$ is the  identification of route of the observed object and its velocity vector prediction.

{\bf C}~--~strategic. The purpose of $\Alg$ is the identification of initial and final points of the trajectory of the object.

On the basis of the provisions it is possible to establish of an integrated system for classification and evaluation of the observed objects, functioning as part of the discussed EOS class.The authors are actively developing the logical-kinematic hypotheses filtration for task~(\ref{eq:Alg_main}) in the discussed statement.

\section{Wide-angle optical systems}
\label{sect:OS}
\setcounter{equation}{0}

Wide-angle optical systems with the individual sensor field of view from 45 x 45 to 180 x 180~degrees are usually used in the discussed class of EOS. The operation of primary detection filter of modern EOS with FPA is based on analysis of the structure-brightness, structure-statistical and geometrical characteristics of the observed objects and background\cite{bib:report_Peli_02}. The geometrical shapes of observable objects formed at the FPA are divided in three classes in compliance with their linear dimensions:
 \begin{itemize}
 \item {\em Full image}~--~the geometry of the object is distorted, but generally retained; the linear dimensions of the image are higher than $11\times11$~pixels;
 \item {\em Pseudoimage}~--~the geometry of the object is weakened, the linear dimensions of the image are from $5\times5$  to $11\times11$~pixels;
 \item {\em Multipoint image}~--~the geometry of the object is singular, the linear dimensions of the image from $1\times1$ to $4\times4$~pixels.
 \end{itemize}

Many direct finder EOS process either “Multipoint image” or “Pseudoimage”; “Full image” for them is an exception. Therefore one of the main tasks is to correctly process the distorted spatial structure of the signal response at the FPA output. The main causes of spatial distortion of the signal are:

\begin{enumerate}
\item The turbulence of propagation medium.
\item The blur caused by the mutual displacement of the observed object and the EOS.
\item The irregular point spread function (PSF) formed by optical system.
\end{enumerate}

In these cases the spatial distortions of the signal response result in the following negative effects:

\begin{enumerate}
\item The energy of the signal response is smeared between the “object” and “background”. It leads to the decreasing of the brightness and statistical contrast and a distortion of the angular size of the object, and, in general, reduces the detectability of the object.

\item The energy in the signal response is redistributed among the “elements of the object”. It leads to the errors in determining the object's coordinates and, in general, reduce the recognizability of the object.
\end{enumerate}

The analysis of the open scientific and technical sources shows that the question of the optical systems optimization using “quality of the PSF” as the criterion is often raised in the aspect of so-called “concentration of energy in a control area”. Some publications devoted to the choice of the optimum size of the area. In this case the homogeneity of the structural characteristics of the PSF by the field of view, including the shape of its peak, is not relevant. However, it is very important in optical systems with fields of view more than 45 degrees\cite{bib:report_PSF_10}.

The second aspect of high-performance EOS wide-angle optical paths design is the uniformity of the instantaneous field of view and illumination in the plane of the receiver over the entire field of view of the system. It allows to get a uniform potential ratio “signal/background” in the case of statistically homogeneous backgrounds and to improve the properties of the EOS as a whole. These two characteristics are closely associated with distortion in wide angle lenses\cite{bib:article_Distorsion_10}.

Infrared systems worked in MWIR and LWIR bands are widespread modifications of EOS. They are usually applied photon detectors working in the BLIP mode (Background Limited Infrared Photodetectors). The authors propose two techniques which allow to designing high-quality optical path of such systems. They are briefly described in the following two subsections.

\subsection{MINOS Technology}
\label{sect:MINOS}
\setcounter{equation}{0}

If the receiver works in BLIP mode, the threshold flux equivalent to the receiver noise is limited by background illumination fluctuations, which also include stray radiation from the warm optical system. In this case stray radiation exerts negative influence on signal-to-background ratio. Thus it's important to minimize stray radiation in the optical system\cite{bib:report_Peli_02}.

For a preliminary assessment of EOS characteristics it is necessary to determine the stray flux at the design stage, and to make changes in construction when needed. As stated in \cite{bib:book_Fischer_08} internal and external stray flux is often assessed in the preproduction model. Therefore their reduction needs considerable efforts and artificial measures that adversely affect the potentially achievable performance of designed systems. But the EOS with extreme specifications are in demand nowadays. Thus a new technology to investigate in detail the issue of internal and external stray illumination is need. Some papers\cite{bib:book_Fischer_08} proposed the use of nonsequential ray tracing to determine the stray flux, but did not describe the mechanism for its implementation. This problem can be solved with new approach proposed by authors: “Technology MINOS”\cite{bib:article_MINOS_09}~--~ab initio method which is free from some limitation inherent to empirical approaches\cite{bib:book_Fischer_08}. It allows to estimate the integral (internal and external) stray flux of infrared (3-5 and 8-14 microns) optical systems at the design stage. In addition, the method shows a way to minimize interfering radiation.

The authors have developed a universal mathematical model which can be customized for the specific structural parameters of the construction and optical system, taking into account the properties of the medium of propagation. The model\cite{bib:article_MINOS_09} based on nonsequential raytracing\cite{bib:book_Fischer_08} from the source (it uses the computational core of the optical CAD) to the detector.

The model operates with the following objects: the sources of radiation, the medium of propagation, radiation detectors. Sources of radiation are:
 \begin{itemize}
 \item optical elements (lenses, mirrors);
 \item elements of mounts;
 \item other structural elements of the optical path;
 \item external sources of stray radiation.
 \end{itemize}
The leading characteristics of the sources and conditions of propagation are:
 \begin{itemize}
 \item a specific design of optical system;
 \item the spatial distribution of temperature in the optical path;
 \item absorption / reflection / scattering coefficients of optical elements materials, the elements of mounts and other structural elements of the optical path;
 \item radiance sources of stray radiation, the parameters of the propagation medium.
 \end{itemize}
The configuration of detectors is determined by the set of parameters:
 \begin{itemize}
 \item format the receiver~--~the number of elements in columns and rows;
 \item the lattice spacing of sensors;
 \item the size of the sensor;
 \item the shape of the photosensitive sensor area;
 \item size of the photosensitive sensor area;
 \item reflection coefficient of the receiver surface.q
 \end{itemize}
The model generates the correct assessment of:
 \begin{itemize}
 \item Absolute (W/m$^2$)values of stray flux incident on the receiver;
 \item the spatial distribution of flux on the surface of the receiver;
 \item the relative contribution of each external source and each element of the optical path to the total stray flux.
 \end{itemize}

Minimization of the integral (internal and external) stray flux of optical systems is provided by solving the inverse problem through: variation of optical properties of the surface of mounts and structural components of the optical system and the optimization of the shape of the inner surfaces of frames.

\subsection{Lens relocation into cold area}
\setcounter{equation}{0}

Currently, cooled detectors are widely used in MWIR band systems. The aperture of such detector is limited by built-in diaphragm (known as cold stop) in the cryogenic area of detector. If the aperture stop of the optical system is situated at cold stop position, the “signal/noise” ratio increases because the detector can see energy only from scene but not from warm mounts (in case of zero vignetting). But the AS position is a “strong” optimization parameter which highly influences on size and quality of the optical system\cite{bib:book_Shannon_97, bib:book_Mann_09}. The best AS position is determined by aberration calculation during the optimization. So there is a contradiction which is acute especially for wide angle systems.

As a possible way of resolving this conflict associated with the position of the AS, the authors propose to transfer some lenses into the cold area. The approach is similar in idea  to the one shown in\cite{bib:report_Singer_10}.

For validation this hypothesis, the authors considered two variants of EOS creation: in the first one the spherical field of view is based on the 6~sensors with an angular field of~90x90 degrees, and in the second~--~on 24~sensors with a field of view of 45x45~degrees. The systems used 1280x1024 and 640x512 pixels MWIR FPA respectively; the pixel size is 15~microns~(which is consistent with the current MWIR systems). This FPA ensure the same IFOV for the systems. The angular fields of view in image space 1024x1024 and 512x512 pixel. In the first case there is a cutoff filter in cold area near photosensitive plane; in the second~--~the filter is combined with a cooled~AS.

The task was to synthesize a high-aperture systems~(f/1) with different field of view and the same picture quality (energy concentration of more than 85~\% in a 2x2~pixel square). The distortion of the systems are close to “f-theta” (deviation from f-theta condition less than 5~\%) to achieve a uniform IFOV over the field\cite{bib:article_Distorsion_10, bib:book_Gerald_04}. Additionally authors tried to decrease size of the system and quantity of the lenses in the scheme while maintaining the same quality. The use two 10-order aspheric surfaces was allowed (parameter~$N_{LA}$).

The elements relocation to a cold area reduced both longitudinal and transverse dimensions of the optical system-receiver assembly. Note that for the system with a field of 45~degrees there is a variant with only one lens outside of cold area; the lens is situated near the entrance window of the detector. The optimization results are shown in Table~\ref{tbl:OS_param} (linear dimensions are in millimeters).
\begin{table}[!htb]
\begin{center}
\caption{Optical systems parameters}\label{tbl:OS_param}
\begin{tabular}{|c|c|c|c|c|c|c|c|c|}
\hline
Variant &$\Omega_{OS}$, degrees &Size &$N_L$/$N_{LC}$ & $N_{LA}$ &$L_{OS}$ &$L_{R}$ &$D_{FL}$ &$D_W$ \\ \hline
1.1 &45 &640x512 &4/0 &2 &60 &36 &20 &26 \\ \hline
1.2 &45 &640x512 &3/2 &1 &41 &14 &22 &13 \\ \hline
2.1 &90 &1280x1024 &6/0 &2 &130 &97 &36 &39 \\ \hline
2.2 &90 &1280x1024 &6/3 &2 &100 &67 &38 &28 \\ \hline
\end{tabular}
\end{center}
\end{table}

As seen from Table~\ref{tbl:OS_param}, the length of the lens~$L_{OS}$ with a field of 90 degrees is~2.7 times greater than that of 45~degree condition, the front lens diameter~$D_{FL}$ is 1.8~times greater, the number of lenses is 6 and 4 respectively~(parameter~$N_L$). Thus, narrow field systems have smaller dimensions~(see also parameter~$L_R$~--~length of the lens to the entrance window of the receiver). It should also be noted that the 640x512 FPA are in the production for a long time, and their manufacturing technology is well established. Matrices with dimensions of 1280x1024~allow us to apply the system with a smaller number of individual sensors, but they require a more complex optical system, they are more expensive, harder to manufacture, and they have greater size.

The lenses location in the cold zone~(parameter~$N_{LC}$) trends to reduce stray flux of the optical system (see Sec.~\ref{sect:MINOS}). This greatly improves “signal/noise” ratio in BLIP mode because the lenses in the warm zone contribute significantly to stray flux.

In addition, the transfer has reduced the diameter of the receiver's window~$D_W$, which has a positive effect on the thermal conditions of the cryogenic area, because of lower heat flow through the germanium window. But the greater amount of lenses amount in the cold zone requires higher cooler power or an increase in the FPA readiness time. The parameters of the cryogenic detector (effective volume) for different variants of the optical system are given in Table~\ref{tbl:R_param}.
\begin{table}[!htb]
\begin{center}
\caption{Cold zone parameters}\label{tbl:R_param}
\begin{tabular}{|c|c|c|c|c|c|c|}
\hline
Variant &Array Format &$N_{LC}$ &$L_{CA}$, mm &$D_W$, mm &$D_{CA}$, mm &$V_{CA}$, $mm^3$ \\ \hline
1.1 &640x512 &0 &23.5 &26 &11 &26~651 \\ \hline
1.2 &640x512 &2 &27 &13 &18 &19~227 \\ \hline
2.1 &1280x1024 &0 &32.8 &39 &22 &98~338 \\ \hline
2.2 &1280x1024 &3 &32.5 &28 &22 &64~119 \\ \hline
\end{tabular}
\end{center}
\end{table}
The Table~\ref{tbl:R_param} shows the changes in: the length from entrance window to the photosensitive plane~$L_{CA}$, the diameters of the entrance window~$D_W$ and the area adjacent to the matrix~$D_{CA}$, and the cryogenic volume~$V_{CA}$ of the receiver.

These data shows that the use of lenses in a cold area complicates the FPA manufacturing process and alignment of the system. There is a need to change the design of the receiver. Requirements for coated lenses located in the cryogenic area became stronger~--~they have to work for a long time under large cyclic fluctuations in temperature and not emit gas in a vacuum.

During the synthesis of such optical systems should also pay attention to the fact that the lenses can degrade the uniformity of illumination in the plane of the receiver (this is necessary to control during the optimization). We are going to study this question carefully in future research.

Thus, the transfer of elements into a cold area leads to systems with smaller dimensions, while maintaining the overall image quality and improving the “signal/noise” ratio. The cost of this are some technical difficulties in the production of receivers, the loss of interchangeability of receivers and the inability to use them in different systems.

\section{Conclusion}
\label{sect:conclusion}

The paper discusses the concept of network-centric conformal staring-sensor EOS with spherical field of view construction. EOS conformality means spatial separation of sensor subsystems in space. Moreover, there is not only spatial distribution of the objects in the group but also sensor's spatial distribution within the object. Network-centricity of distributed EOS implies two interrelated aspects: the conformal processing of information from sensor matrix in a single event coordinate-time field and the high level of information completeness with the nonstationarity (mobility) of the matrix.

Abstract passive distributed electro-optical systems with FPA based on a group of moving objects distributed in space are discussed. During the paradigm of such systems interpretation synthesis, the authors based on synergy and cybernetics. It can be explained by two factors: first~--~the distributed network-centric system is measuring-information system, and second~--~it is heterogeneous system (the system includes the human operator).

The report touched upon one of the central problems in the synthesis of network-centric EOS~--~the optimal distribution of functions between the EOS and information-management system. The importance of this issue is caused by ultra-high information integration of network-centric EOS into global information management grid. The authors proposed to find the solution by optimization method, based on an iterative process for the hierarchical complex systems synthesis. This approach emphasizes to the first place comprehensive simulation and modeling technologies that operate under “multiphysics” conception.

The issue of boundary EOS configurations, in which the property of their distribution and network-centricity still remains, is also noted.

The task of logical-kinematic filter synthesis is shown.  This is the authors' approach which is in active development. It aims to restore the fullness of observed objects coordinates without constructing a direct finder triangle. The algorithms allow to restore the most probable trajectory of the observed object in global coordinate system, to classify the object and diagnose its status. The self-consistent nonlinear filter used to obtain these estimates is a feature of the proposed solution. Additionally classification/diagnosis of the object and its trajectory synthesis is carried out jointly. Kinematic trajectory-evaluation of the object's coordinates; logical-probabilistic models of space and environment surrounding the object, the probabilistic and dynamic behavior of the object are combined in the whole. In addition, the algorithm makes a decision based on a priori and operational defined constraints and preferences provided in the form of non-rigid probability inequalities.

A great part of the paper is devoted to the problems of the optical systems for such EOS class synthesis. Wide angle optical systems with an individual sensor field of view from 45x45 to 180x180 degrees are usually used. Therefore the authors pay attentions to fundamental questions for ensuring: the PSF quality over the field; the uniformity of the instantaneous field of view and illumination in the plane of the receiver over the field of view.

Infrared systems worked in MWIR and LWIR bands are widespread modifications of EOS. They are usually applied photon detectors working in the BLIP mode. The authors propose two techniques which allow to design high-quality optical path of such systems.

One of them (MINOS) allows us to estimate the integral (internal and external) of the stray flux in optical systems at the design stage and points to ways to minimize interfering radiation. Technology MINOS~--~an author's development~--~ab initio method, is free from a number of disadvantages of empirical methods. It is based on direct rays tracing from the source (it uses the computational core of the optical CAD) to the detector. A mathematical model adjusts for the specific structural parameters of the test circuit and the optical path and accounts for the properties of the medium of propagation.

The elements relocation to a FPA cold area is a second technology which allows to reduce both longitudinal and transverse dimensions of the optical system-receiver assembly. Such configuration corresponds to a system with lower own radiation, that greatly improves the “signal/noise” ratio for BLIP mode. Additionally the elements relocation decreases the diameter of input optical window and the cryogenic area volume which has a positive effect on the thermal conditions of the cryogenic area. But the greater number of lenses in the cold zone requires higher cooler power or an increase in the FPA readiness time. Requirements for coated lenses located in the cryogenic area became stronger~--~they have to work for a long time under large cyclic fluctuations in temperature and not emit gas in a vacuum. But there are some technical difficulties in the production of receivers, the loss of interchangeability of receivers and the inability to use them in different systems.

Thus, as mentioned above the paper discusses the main problems of network-centric conformal staring-sensor EOS with spherical field of view synthesis. A draft of possible solutions is presented for some of these systems. A detailed study of these issues is the subject of our future publications.

\begin{Biblioen}

\bibitem{bib:article_Cebrowski_98}
\textit{Cebrowski~A.~K., Garstka~J.~H.,} Network-Centric Warfare –- Its Origin and Future, U.S. Naval Institute Proceedings Magazine, {\bf 124}/1/1, 139 (1998).

\bibitem{bib:book_Viner_61}
\textit{Wiener~N.,} Cybernetics: Or Control and Communication in the Animal and the Machine, Paris, (Hermann and Cie) and Camb. Mass, (MIT Press), 2nd revised ed., (1961).

\bibitem{bib:book_Haken_04}
\textit{Haken~H.,} Synergetics: Introduction and Advanced Topics, Springer, (2004).

\bibitem{bib:book_Haken_06}
\textit{Haken~H.,} Information and self-organization: a macroscopic approach to complex systems, Springer, (2006).

\bibitem{bib:article_Storn_97}
\textit{Price~K., Storn~R.,} Differential Evolution -- A Simple and Efficient Heuristic for Global Optimization over Continuous Spaces, Journal of Global Optimization, {\bf 11}, 341 (1997).

\bibitem{bib:book_Storn_05}
\textit{Price~K., Storn~R., Lampinen,~J.,} Differential Evolution: A Practical Approach to Global Optimization, Springer, (2005).

\bibitem{bib:report_Peli_02}
\textit{Peli~T., Monsen~P., Stahl~R., McCamey~K.,} Signal Processing Improvements for Infrared Missile Warning Sensor, Aerospace and Electronics Conference  1997; NAECON 1997; Proceedings of the IEEE 1997 National, {\bf 2}, 1052 (2002).

\bibitem{bib:article_Cardillo_02}
\textit{Cardillo~G.~P., Mrstik~A.~V., Plambeck~T.,} A track filter for reentry objects with uncertain drag, IEEE Trans. on Aerospace and Electronic Systems, {\bf 35}:2,  394 (2002).

\bibitem{bib:article_Daum_03}
\textit{Daum~F., Fitzgerald~R.,} Decoupled Kalman filters for phased array radar tracking, IEEE Trans. on Automatic Control, {\bf 28}:3, 269 (2003).

\bibitem{bib:report_Tartakovsky_00}
\textit{Tartakovsky~A.~G., Blazek~R.~B.,} Effective Adaptive Spatial-Temporal Technique for Clutter Rejection in IRST, SPIE Proceedings: Signal and Data Processing of Small Targets, {\bf 4048}, 85 (2000).

\bibitem{bib:report_Madyastha_05}
\textit{Madyastha~V.~K., Calise~A.~J.,} An Adaptive Filtering Approach to Target Tracking, American Control Conference, Portland, OR, USA, 1269 (2005).

\bibitem{bib:report_WGS84_00}
NIMA Technical Report TR8350.2, \textit{Department of Defense World Geodetic System 1984}, Its Definition and Relationships With Local Geodetic Systems, Third Edition; AMENDMENT 1; 3~January~2000.

\bibitem{bib:report_PSF_10}
\textit{Erokhin~E.~V., Makarenko~A.~V., Pravdivtsev~A.~V.,} About the necessity of PSF evaluation in IR direct-finding stations, International Conference “Digital Signal Processing and Application”, Moscow, Russia,  175 (2010).

\bibitem{bib:article_Distorsion_10}
\textit{Makarenko~A.~V., Pravdivtsev~A.~V., Udin~A.~N.,} The formation constant instanceous field of view in wide angle optical systems, XXI International Conference “Photoelectronics and night vision devices”, Moscow, Russia, 175 (2010).

\bibitem{bib:book_Fischer_08}
\textit{Fischer~R.~E., Tadic-Galeb~B., Yoder~P.~R.,} Optical system design, McGraw-Hill, (2008).

\bibitem{bib:article_MINOS_09}
\textit{Makarenko~A.~V., Pravdivtsev~A.~V., Udin~A.~N.,} The  estimation method of internal stray  radiating in infrared systems, Electromagnetic waves and electronic systems, {\bf 12}, 28 (2009).

\bibitem{bib:book_Shannon_97}
\textit{Shannon~R.~R.,} The art and science of optical design, Cambridge University press, (1997).

\bibitem{bib:book_Mann_09}
\textit{Mann~A.,} Infrared optics and zoom lenses, SPIE Press; Bellingham, (2009).

\bibitem{bib:report_Singer_10}
\textit{Singer~M., Oster~D.,} Design of a Cryogenic IR Detector with Integrated Optics, Infrared Technology and Applications XXXVI, edited by Bjorn~F.~Andresen, Gabor~F.~Fulop, Paul~R.~Norton, Proc. of SPIE, {\bf 7660} (76601Z), 1 (2010).

\bibitem{bib:book_Gerald_04}
\textit{Dekker~M.,} ed. Gerald~F., Handbook of Optical and Laser Scanning, Marshall Inc., (2004).

\end{Biblioen}

\noindent
{\bf Andrey V. Makarenko} -- was born in 1977, since 2002 -- Ph.~D. of Cybernetics. Founder and leader Research \& Development group “Constructive Cybernetics”. Author and coauthor of more than 50 scientific articles and reports. Associate Member IEEE (IEEE Systems, Man, and Cybernetics Society Membership). Research interests: analysis of the structure dynamic processes, predictability; detection, classification and diagnosis is not fully observed objects (patterns); synchronization in nonlinear and chaotic systems; system analysis and modeling of economic, financial, social and bio-physical systems and processes; system approach to development, testing and diagnostics of complex information-management systems.
\\
\\
{\bf Andrey V. Pravdivtsev} -- was born in 1983. In 2010 finished postgraduate study in Bauman Moscow State Technical University, chair “Laser and opto-electronic systems”, speciality “Optic and opto-electronic devices and systems”. Chief specialist in Research \& Development group “Constructive Cybernetics”. Science interest: methods of synthesis and optimization of opto-electronic systems and hyperspectral optical systems. Individual member of Optical Society of
America, member of SPIE.

\end{document}